# Unification of the Pressure and Composition Dependence of Superconductivity in Ru substituted BaFe$_2$As$_2$

T. R. Devidas, Awadhesh Mani[†], Shilpam Sharma, K. Vinod, A. Bharathi and C. S. Sundar

Materials Science Group,
Indira Gandhi Centre for Atomic Research,
Kalpakkam-603102, Tamil Nadu, INDIA

**Abstract:**

Temperature dependent high pressure electrical resistivity studies has been carried out on Ba(Fe$_{1-x}$Ru$_x$)$_2$As$_2$ single crystals with $x = 0.12$, $0.26$ and $0.35$, which correspond to under doped, optimally doped and over doped composition regimes respectively. The evolution of structural/magnetic ($T_{S-M}$) and superconducting transition ($T_c$) temperatures, with pressure for various compositions have been obtained. The normal state resistivity has been analyzed in terms of a model that incorporates both spin fluctuations and the opening of the gap in the spin density wave (SDW) phase. It is shown that $T_c$ scales with the strength of the spin fluctuation, $B$, and $T_{S-M}$ scales with the SDW gap parameter, $\Delta$. This provides a prescription for the unification of the composition and pressure induced superconductivity in BaFe$_2$As$_2$.

---

[†] Corresponding author's e-mail: mani@igcar.gov.in

## INTRODUCTION

Among the various classes of FeAs family, $BaFe_2As_2$ is most widely studied compound as the single crystals of this system can be reproducibly grown relatively easily as compared to other arsenides [1]. $BaFe_2As_2$ exhibits SDW ordering of Fe spins in concomitance with the structural transition from the tetragonal to an orthorhombic phase below $T_{S-M}$ ~ 132-140K [1-6]. Various chemical substitutions involving hole doping in $Ba_{1-x}K_xFe_2As_2$ [3], electron doping in $Ba(Fe_{1-x}Co_x)_2As_2$ [4] and isovalent doping in $Ba(Fe_{1-x}Ru_x)_2As_2$ and $BaFe_2(As_{1-x}P_x)_2$ systems [5-6] have been investigated. These studies reveal broadly similar features of temperature – composition ($T - x$) phase diagrams in which doping progressively suppresses the structural-magnetic transition temperature ($T_{S-M}$) of the parent compound in favour of superconductivity near a critical composition $x_c$. The superconducting transition temperature ($T_c$) exhibits dome like feature with $x$ demarcating three composition regimes: (i) under doped where $T_c$ increases with $x$, (ii) optimally doped regime around the composition where $T_c$ reaches to a maximum value and (iii) over doped regime in which $T_c$ falls with $x$. Besides chemical substitutions, the external pressure has been extensively used as another valuable tool to induce superconductivity in pristine as well as doped $BaFe_2As_2$ system [2, 7-11].

It has been of interest to see if the pressure and composition induced superconductivity in the Iron-arsenides have a common basis. The equivalence of chemical and external pressure effects in dictating the evolution of $T_c$ has been investigated in isovalent substituted systems such as $BaFe_2(As_{1-x}P_x)_2$ and $Ba(Fe_{1-x}Ru_x)_2As_2$ [10-11]. In $Ba(Fe_{1-x}Ru_x)_2As_2$ system, in which Ru substitution increases the lattice volume, high pressure studies in under doped and optimal doped regimes revealed that $T_c$ exhibit an increase followed by decrease with pressure [11]. Interestingly, the $T-P$ phase diagrams of the several compositions of Ru studied could be collapsed into a single *T-P* phase diagram by shifting the origin of the pressure axis by 3 GPa for 10% increase in Ru [11]. This however cannot be rationalized in terms of the associated structural changes, since high pressure is seen to decrease all lattice parameters (*a*, *c*) and lattice volume V [8], whereas Ru substitution increases *a* while decreases *c* giving rise to an increase in V [4].

Here, we investigate the equivalence of pressure and composition in inducing superconductivity in the Ba(Fe$_{1-x}$Ru$_x$)$_2$As$_2$ system through an analysis of the strength of the spin fluctuations, rather than any structural parameters. The strength of the spin fluctuations has been evaluated from an analysis of the normal state resistivity. Experiments on the temperature dependence of resistivity under pressure across the SDW and superconducting transition have been carried out in Ba(Fe$_{1-x}$Ru$_x$)$_2$As$_2$ single crystals with $x = 0.12$, 0.26 and 0.35, which correspond to under doped, optimally doped and over doped composition regimes respectively. It is shown that $T_c$ scales with the strength of the spin fluctuation, *B*, and the structural/magnetic transition, $T_{S-M}$, scales with the SDW gap parameter, $\Delta$.

**EXPERIMENTAL DETAILS**

Single crystal samples of Ba(Fe$_{2-x}$Ru$_x$)$_2$As$_2$ with nominal composition of $0 \leq x \leq 0.4$ were prepared by slow cooling a stoichiometric mixture of Ba chunk, FeAs and RuAs powders without using any flux [12]. The powder x-ray diffraction (XRD) measurements on ground-single crystals of all the nominal compositions were performed to characterize the phase formation, to identify crystal structure and to determine the lattice parameters by using a STOE diffractometer operating in the Bragg-Brentano geometry [13]. High pressure resistivity measurements as a function of temperature were carried out on a single crystalline sample of Ba(Fe$_{1-x}$Ru$_x$)$_2$As$_2$ ( $0 \leq x \leq 0.35$) by mounting only a single piece of a given composition in an opposed anvil pressure locked cell following the procedure described in ref. [7].

**RESULTS & DISCUSSIONS**

The $T$ - $x$ phase diagram of Ba(Fe$_{1-x}$Ru$_x$)$_2$As$_2$ ( $0 \leq x \leq 0.35$) single crystals exhibiting the variation of $T_{S-M}$ and $T_c$ as a function of Ru content ($x$), deduced from the temperature dependent resistivity ρ(T) measurements at ambient pressure in 4 to 300 K range [13], is depicted in Fig. 1(a). $T_{S-M}$ progressively shifts towards lower temperature from 139 K for x = 0 to 52 K for $x$ = 0.21 and vanishes beyond this concentration. The superconductivity appears for $x$ = 0.14 with $T_c$ ~ 7 K, then reaches to a maximum value of $T_{c,max}$ = 21 K for $x$ = 0.26 and finally decreases to 20 K for $x$ = 0.35 ( see Fig. 1(a)). From the figure it is clear that the samples $x$ = 0.12, 0.26 and 0.35 (marked by arrows) belong to under doped, optimally doped and over Ru-doped regimes respectively, and they have been chosen for the present high pressure electrical resistivity studies.

In Figs. 1(b)-(g), we show the evolution of $T_{S-M}$ and $T_c$ with pressure for Ba(Fe$_{1-x}$Ru$_x$)$_2$As$_2$ with x =0.0, 0.12, 0.26, & 0.35, as obtained from high pressure low temperature resistivity measurements. The $T – P$ data for the undoped sample (x = 0), shown in Fig 1 (b), has been included here from our previous high pressure studies for comparison [7]. The pressure dependence of $T_c$ for various compositions is shown in the right panel, with the same scale to enable better comparison (cf. figs. 1(d) – (g)). It is seen from Figs. 1(b) and 1(c) that the $T$-$P$ phase diagrams of pristine BaFe$_2$As$_2$ and under-doped ($x$ = 0.12) samples are qualitatively similar. In the $x$ = 0 sample, $T_{S-M}$ monotonically decreases with pressure and beyond $P$~ 1.5 GPa its signature could not be discerned from ρ(T) curve. The maximum superconducting transition temperature $T_{c, max}$ ~ 35.4 K occurs around 1.5 GPa pressure. For $x$ = 0.12 sample, $T_{S-M}$ decreases from 103 K at $P$ = 0 GPa to 48 K at $P$ ~ 0.8 GPa and vanishes beyond this pressure. Superconductivity appears at $P$ ~ 0.4 GPa, attains $T_{c,max}$ = 34.7 K at $P$ = 0.8 GPa and then decreases beyond this pressure. The optimally Ru-doped sample ($x$ = 0.26) exhibits superconductivity at ambient pressure with $T_c$ = 21 K, without any noticeable signature of $T_{S-M}$. The variation of its $T_c$ versus $P$, depicted in Fig. 1 (f), shows non-monotonic variation similar to that of the undoped and under doped samples. However, in the case of over-doped sample, $x$ = 0.35, the $T_{c,max}$ = 20.5 K occurs at ambient pressure itself, and $T_c$ shows a sharp ($dT_c/dP$ ~ -10 K/GPa ) reduction with pressure vanishing at ~ 2GPa. The vertical dotted line in Fig. 1 (right panel) clearly indicate that the pressure ($P_{max}$) at which maximum $T_{c,max}$ occurs, shifts to lower values with increasing $x$. This implies that effect of Ru substitution is equivalent to applying the external pressure as far as occurrence of $T_{c,max}$ is concerned.

It is widely appreciated [6-9] that the control parameters ($P$ and $x$) drive the $BaFe_2As_2$ system through a magnetic quantum critical point (QCP) at critical $x_c$ or $P_c$, where $T_{S-M}$ tends to zero while an asymmetric superconducting dome emerges with $T_{c,max}$ occurring in the vicinity of QCP. Theoretical studies suggest that superconductivity in vicinity of QCP is associated with the spin fluctuations [14]. Experimental studies, such as NMR and neutron scattering [15-16], implicate the role of spin fluctuations in the superconductivity of FeAs system. Therefore, the strength of spin fluctuations appears to be an appropriate physical parameter in unifying the dependence of $T_c$ with pressure and composition. In what follows, we try to rationalize the observed T-x and T-P phase diagrams of $Ba(Fe_{1-x}Ru_x)_2As_2$ in terms of the strength of the spin fluctuations.

The normal state ρ(T) of FeAs- system is dominated by AFM spin fluctuation in $T > T_{S-M}$ temperature regime [7, 9], while spin density wave (SDW) transition below $T_{S-M}$ opens a spin gap (Δ) in the AFM ordered state. An appropriate expression for the ρ(T) which takes into account of the above contributions, in addition to a general contribution due to phonon scattering can be given as [7, 9, 17]:

$$\rho = A + BT^m + CT(1 + 2T/\Delta)\exp(-\Delta/T) + D\left(\frac{T}{\Theta}\right)^5 \int_0^{\Theta/T} \left[\frac{x^5 dx}{(e^x - 1)(1 - e^{-x})}\right] \text{--------- (1)}$$

The ρ(T) data is fitted in the two temperature regimes as follows: In $T > T_{S-M}$ regime, where m=3/2 is taken to account for resistivity contribution of the AFM spin fluctuation scattering [7, 8], all the ρ(T) data are first plotted as a function of $T^{3/2}$. Based on the observed linearity of the plot, the parameters *A* & *B* are obtained from the linear fit. Parameter *A* represents electron-impurity scattering contribution to the resistivity, while *B* is a measure of strength of AFM fluctuation [7]. In some cases, as described in footnote [18], the ρ(T) versus $T^{3/2}$ plots were seen to exhibit a minor deviation from the exact linearity. In such cases, the last term of Eq. (1) representing electron-phonon interaction is also taken into account to get the best fit. The third term with coefficient *C* arising due to AFM ordering below $T_{S-M}$ is not applicable in T>$T_{S-M}$ regime. In $T_c$<T<$T_{S-M}$ regime, we took m=2 to account for the Fermi-liquid contribution and the third term with coefficient *C* and analyze the ρ(T) data to deduce AFM spin energy gap (Δ).

The left panel of Fig. 2 shows fittings of ambient pressure ρ(T) data of Ba(Fe$_{1-x}$Ru$_x$)$_2$As$_2$ samples with (0 ≤ *x* ≤ 0.35) in T>$T_{S-M}$ regime. A goodness of the fit to $T^{3/2}$ law indicates the dominance of AFM fluctuation over electron-phonon scattering process in dictating the ρ(T) behaviour of this system above $T_{S-M}$. The upper inset of Fig. 2 (left panel) shows a systematic decrease in parameter $A$ with increase in $x$. This could arise because of the addition of carriers by Ru doping (see Ref. [12]), resulting in an overall decrease of resistivity (see Fig 2), over and above the impurity scattering effect. This happens from the fact that Ru substitution which dominates over the contribution due to impurity scattering. The inset to figure shows the variation of $B$ and $T_c$ with $x$. It is also seen that the range of the $T^{3/2}$ fits extends to lower temperatures with the increase of Ru content, consistent with the decrease of $T_{S-M}$, seen in Fig. 1(a). The right panel of Fig. 2 depicts fitting for representative ρ(T) of Ba(Fe$_{1-x}$Ru$_x$)$_2$As$_2$ samples in $T_c<T<T_{S-M}$ temperature regime. The value of Δ extracted from the fit is found to be 130 K for $x = 0$, which is in a good agreement with value of 9.8 meV (~115K) obtained from inelastic neutron scattering studies [19]. The variation of Δ with x is shown in the inset of Fig. 2 (right panel). Δ monotonically decreases in similar fashion as $T_{S-M}$ with increasing x and vanishes beyond $x = 0.21$.

The pressure dependent normal state ρ(T, P) data of samples with $x = 0.12$, 0.26, and 0.35 have been analyzed using Eq. (1) in temperature regimes $T>T_{S-M}$ and $T<T_{S-M}$ using the aforementioned procedures. The representative ρ(T) data at high pressures for the x=0.12 sample is shown in Fig. 3 (a). The variation of Δ with pressure, as obtained from the analysis of resistivity data, for x=0.12 is shown in the inset of Fig. 3 (a). It is seen that Δ decreases with increasing $P$ and vanishes beyond $P\sim 0.8$ GPa. The lower inset shows variation of $A$ and $D$ with $P$. Parameter $A$ decreases, while $D$ marginally increases with pressure. The pressure dependence of parameter $B$ for $x = 0$, 0.12, 0.26 and 0.35 is presented in Figs. 3 (b) - (e) respectively. The pressure dependence of $T_c$ of the corresponding systems (cf. Figs.1(d)-(g)) is also shown on the right ordinate. A striking correlation in the variation of $B$ and $T_c$ can be clearly seen from these figures. For the undoped ($x = 0$), under doped ($x = 0.12$) and optimally doped ($x = 0.26$) samples (Figs. 3 (b)-(d)), $B$ and $T_c$ both exhibit non-monotonic variation with pressure, while for over-doped ($x = 0.35$) sample $B$ and $T_c$ decrease monotonically with $P$ (Fig. 3e). This correlation in $B$ and $T_c$ indicates that the strength of AFM spin fluctuations is a crucial parameter intimately related with the evolution of superconductivity in $BaFe_2As_2$ system. A similar tracking of spin fluctuations, obtained from spin-lattice relaxation rate of the normal state in high pressure NMR measurements, with the measured $T_c$ has been recently reported in $NaFe_{0.94}Co_{0.06}As$ system [20].

In Fig. 4, we consolidate the results of both the compositional and pressure dependence of $T_{S-M}$ and $T_c$ in $Ba(Fe_{1-x}Ru_x)_2As_2$, along with the results of resistivity analysis, leading to the extraction of SDW gap parameter $\Delta$ and the spin-fluctuation parameter $B$ ( see Eq. 1 ). In Fig. 4(a) we have plotted all the $T_{S-M}$ versus $\Delta$ data, obtained from the analysis of Ru substituted $\rho(T)$ as well as pressure dependent $\rho(T, P)$ data of $x = 0.12$, which shows a linear relationship. The AFM ordering temperature $T_{S-M}$ is predicted to vanish near the critical parameter $P_c$ (in present case - composition $x_c$ / pressure $P_c$) following the expression [21];

$$T_{S-M} \alpha \ |P - P_c|^\alpha \text{------------------ (2)}$$

where $\alpha$ is critical exponent with a value of 1 and 2/3 for the two and three dimensional AFM systems respectively. A fit to the $T_{S-M}$ versus $x$ data of Fig 1(a) to Eq. (2) yields the critical Ru content $x_c \sim 0.25\pm0.02$ and $\alpha = 0.62\pm0.03$. The $x_c$ thus obtained is the optimally doped composition. Similarly, the fit to $T_{S-M}$ versus $P$ data of x = 0 (and 0.12) (Figs. 1(b) - (c)) give rise to $P_c \sim 1.95\pm0.05$ GPa (1.15±0.05 GPa) and $\alpha = 0.54\pm0.07$ (0.70±0.03). The values of exponent thus deduced are closer to $\alpha \sim 2/3$ rather than 1 suggesting three dimensional (3-

D) nature of AFM spin fluctuations in this system in agreement with the results from inelastic neutron scattering and angle resolved photoemission spectroscopy [22-23].

The bottom panel of Fig. 4(b) shows $T_c$ versus $B$ plot obtained from all the Ru substituted and pressure dependent data in Ba(Fe$_{1-x}$Ru$_x$)$_2$As$_2$. Strikingly, it is seen that all the composition and pressure dependence $T_c$ vs. $B$ data, fall on a single curve except the high pressure data of $x$=0. Even for the undoped sample ($x = 0$), $T_c$ follows $B$ initially up to $P_c \sim$ 2GPa beyond which $T_c$ gradually decreases with increase of $B$. It is seen from Fig. 4(b) that the variation of $T_c$ with $B$ gets segregate into two regimes. In first regime with moderate value of $B$ ($\sim 10^{-4}$ $\mu\Omega$-m-K$^{-3/2}$), $T_c$ increase with increasing $B$, while in second regime with larger value of $B$ ($\sim 10^{-3}$ $\mu\Omega$-m-K$^{-3/2}$) $T_c$ decreases monotonically with $B$. This indicates that there may exist a critical value of spin fluctuation strength $B_c$ beyond which spin fluctuations is detrimental to superconductivity. If so, then the linear extrapolations of $T_c$ versus $B$ plot from both the regimes intersecting at $B_c \sim$ ( 4.9x10$^{-4}$ $\mu\Omega$-m-K$^{-3/2}$ ) gives a $T_c$ of $\sim$ 42 K, which would be the maximum achievable $T_c$ in BaFe$_2$As$_2$ system. Interestingly, the maximum $T_c$ of ~38 K has been seen in K–doped BaFe$_2$As$_2$ system [3]. It should be remarked that theoretical study [24] involving spin fluctuation mediated pairing also reveals the similar non-monotonic evolution of $T_c$ as a function of coupling strength ($u$), i.e., $T_c$ initially increases with $u$ for weak coupling, passes through a maximum at intermediate coupling and then decreases for strong coupling strength.

**SUMMARY**

In the present study on Ba(Fe$_{1-x}$Ru$_x$)$_2$As$_2$ system, we have measured the $T$-$x$ phase diagram and the $T$-$P$ phase diagrams for various compositions, spanning the under doped, optimally doped and over-doped regimes. The compositional and pressure dependence of $T_c$ for various *x*, can be unified, if viewed in terms of the underlying strength of the spin-fluctuations – that has been obtained from the analysis of normal state resistivity.  It would be of interest to see if the correlation as seen in Fig. 4(b), is seen for other doped systems.

**Figure 1.**

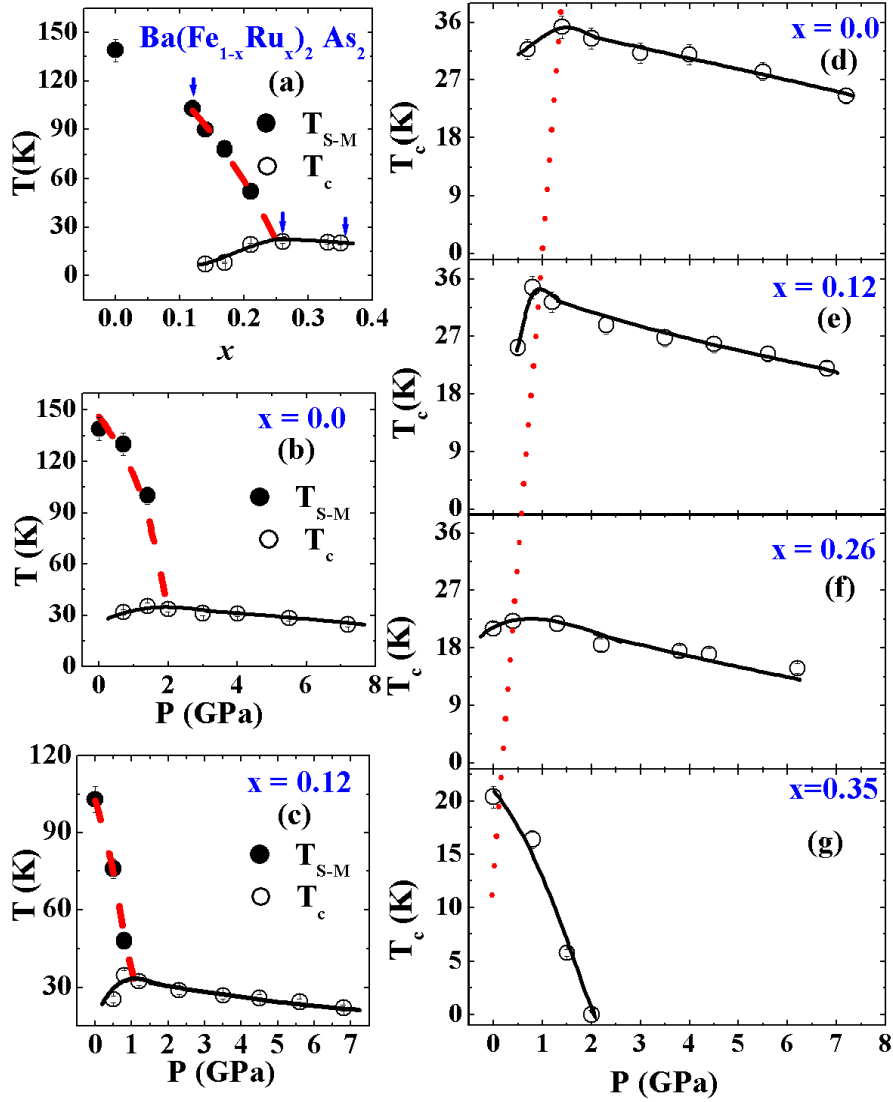

**Fig. 1 (a)** Variation of $T_c$ and $T_{S-M}$ as a function of Ru content (x) depicting *T-x* phase diagram for Ba(Fe$_{1-x}$Ru$_x$)$_2$As$_2$ (Ref. 13). **(b)-(c)** Variation of $T_{S-M}$ & $T_c$ as a function of pressure for x=0 and x=0.12 respectively. **(d)-(g)** $T_c$ versus pressure plots for x=0, 0.12, 0.26 and 0.35 respectively. The dash-dot line (left panel) joining $T_{S-M}$ in **(a)** –**(c)** are fit to Eq. 2. Dash line (right panel) joining the $T_{c,max}$ of various *x* shows shifting of $P_{max}$ towards lower pressure with increase in x. Rest of the solid lines are guides to eyes.

**Figure 2.**

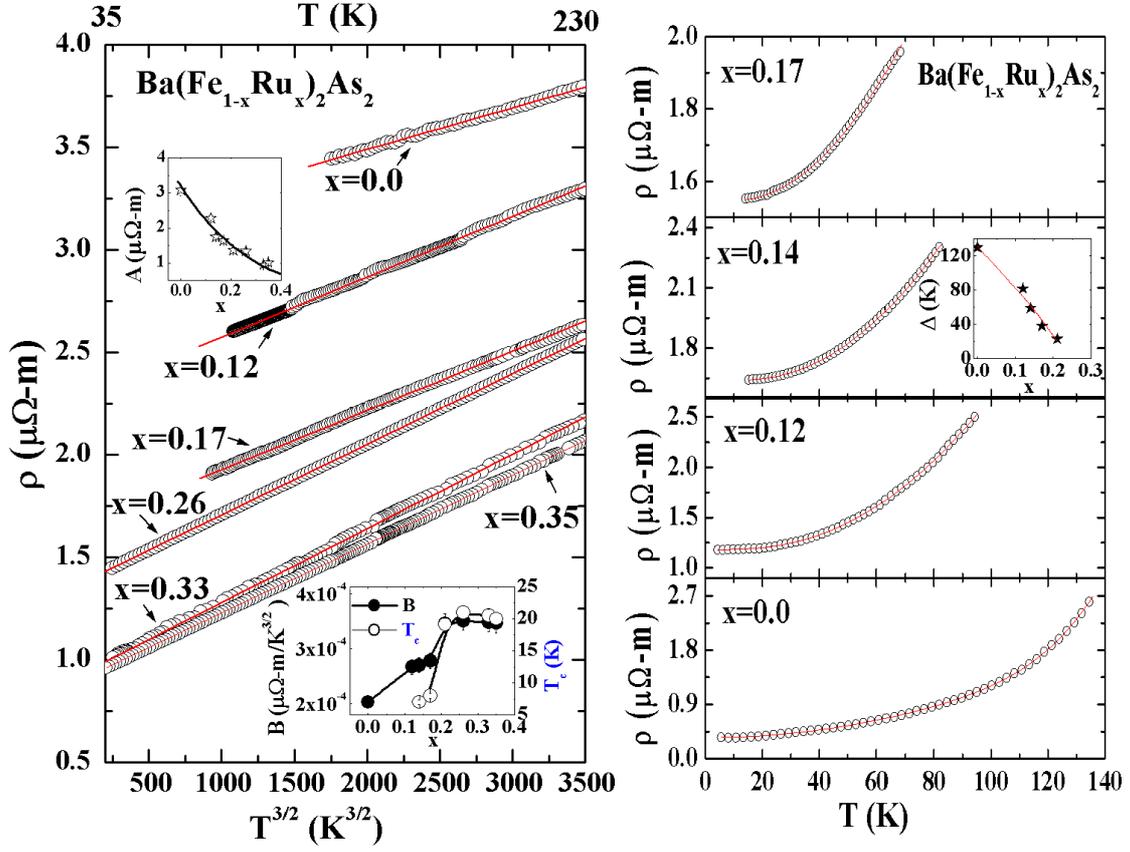

**Fig. 2:** (Left panel) ρ versus $T^{3/2}$ plot for various Ba(Fe$_{2-x}$Ru$_x$)$_2$As$_2$ sample showing fit to Eq. 1 for T > T$_{S-M}$. The variations of *A* (upper inset), *B* and T$_c$ (lower inset) are shown as a function of x. (Right panel) Fitting of ρ(T) data to Eq. 1 for T$_c$ < T < T$_{S-M}$. The variation of Δ as a function of Ru content x is shown in the inset of this panel.

**Figure 3.**

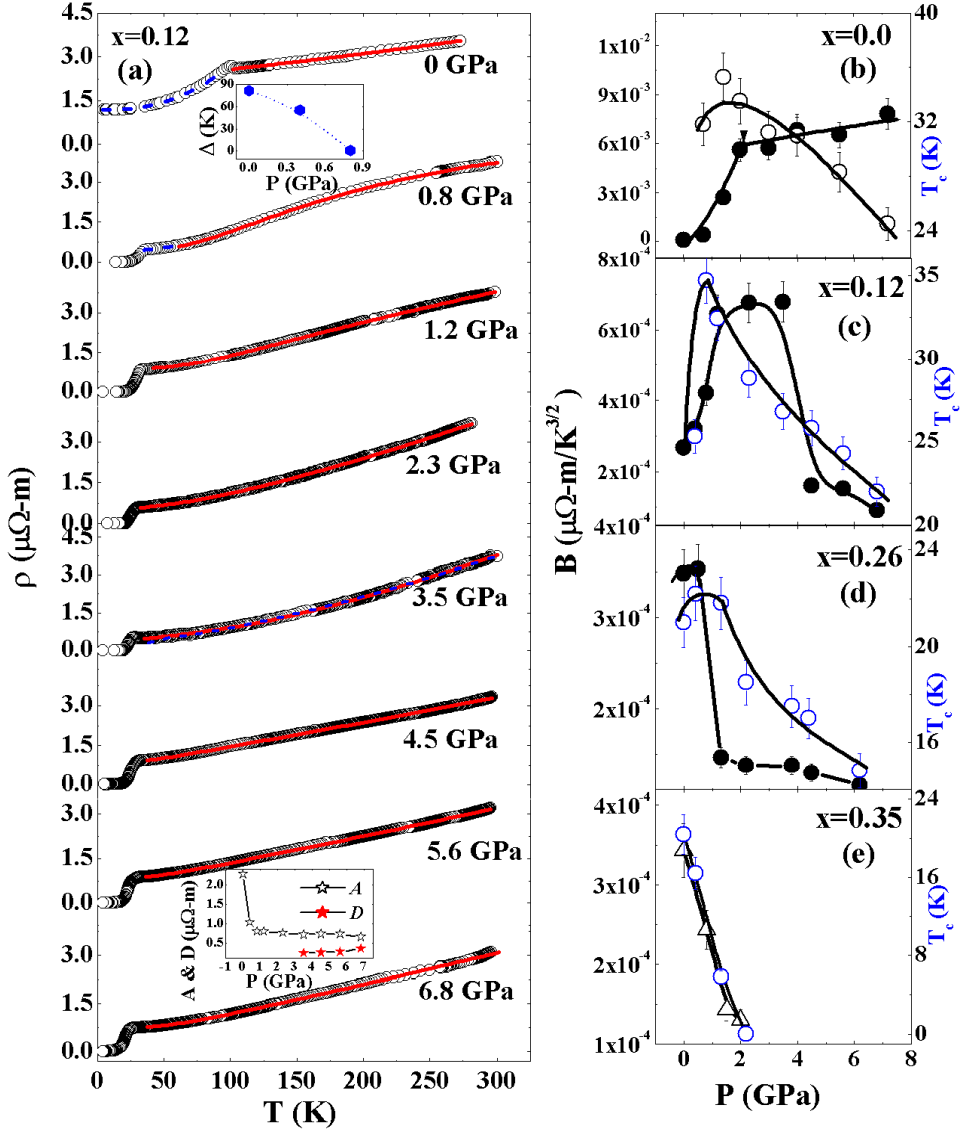

**Fig. 3 (a)** $\rho(T)$ of $Ba(Fe_{1-x}Ru_x)_2As_2$ with x=0.12 at various representative pressures between 0 and 6.8 GPa showing fit to Eq. 1 for $T>T_{S-M}$ (solid red line) and $T<T_{S-M}$ (dash blue line). The variations of $\Delta$ (upper inset), $A$ and $D$ (lower inset) are shown as a function of P. Fits with $D$ term (red solid line) and without $D$ term (green dash line) for $P$=3.5 GPa curve is shown for comparison. **(b)-(e)** exhibits variation of $B$ with pressure $P$ for x=0.0, 0.12, 0.26 and 0.35 respectively.

**Figure 4.**

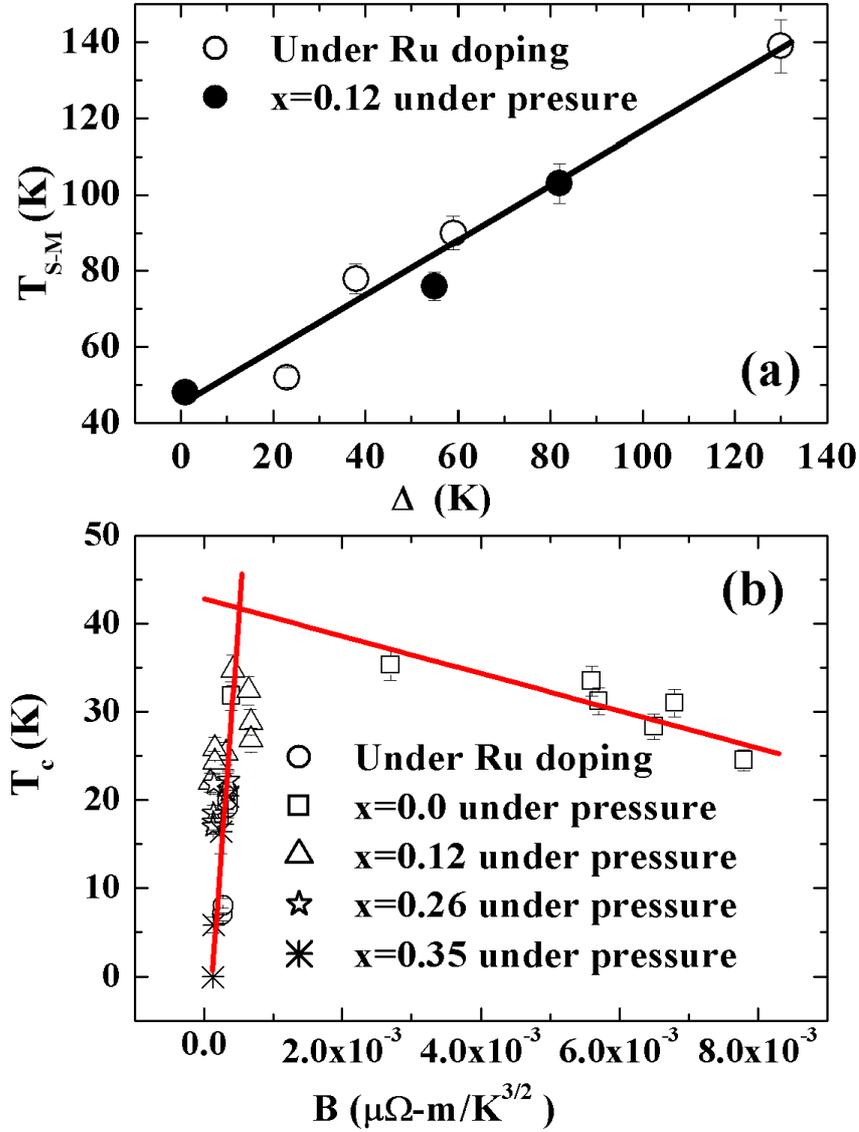

**Fig. 4** (a) Correlation in $T_{S-M}$ and $\Delta$ obtained from Ru doped system and x=0.12 sample under pressure. (b) $T_c$ versus parameter B plot obtained from all data, i.e., Ru substituted samples as well as for samples x=0, 0.12, 0.26 and 0.35 under pressure.